\documentclass[letter]{aa} 

\usepackage{hyperref}

\usepackage{amsmath}
\usepackage{graphicx}
\usepackage{graphics}

\usepackage{natbib}{}
\usepackage{xcolor}
\usepackage{url}

\bibpunct{(}{)}{;}{a}{}{,}

\def\teff{\ifmmode T_{\rm eff} \else $T_{\mathrm{eff}}$\fi}

\def\ltsima{$\buildrel<\over\sim$}
\def\lsim{\lower.5ex\hbox{\ltsima}}

\newcommand{\hii}{H~{\sc ii}}
\newcommand{\ha}{\ifmmode {\rm H}\alpha \else H$\alpha$\fi}
\newcommand{\hb}{\ifmmode {\rm H}\beta \else H$\beta$\fi}
\newcommand{\hg}{\ifmmode {\rm H}\gamma \else H$\gamma$\fi}
\newcommand{\lya}{\ifmmode {\rm Ly}\alpha \else Ly$\alpha$\fi}
\newcommand{\hei}{He~{\sc i}}

\newcommand{\heii}{He~{\sc ii}}

\newcommand{\Heiiopt}{He~{\sc ii} $\lambda$4686}

\newcommand{\ebv}{\ifmmode E_{\rm B-V} \else $E_{\rm B-V}$\fi}
\newcommand{\av}{\ifmmode A_{\rm V} \else $A_{\rm V}$\fi}

\def\micron{$\mu$m}
\def\kms{km s$^{-1}$}

\def\cmc{cm$^{-3}$}

\def\msun{\ifmmode M_{\odot} \else M$_{\odot}$\fi}
\def\msunyr{\ifmmode M_{\odot} {\rm yr}^{-1} \else M$_{\odot}$ yr$^{-1}$\fi}
\def\zsun{\ifmmode Z_{\odot} \else Z$_{\odot}$\fi}

\def\lsun{\ifmmode L_{\odot} \else L$_{\odot}$\fi}

\def\mup{\ifmmode M_{\rm up} \else M$_{\rm up}$\fi}
\def\mlow{\ifmmode M_{\rm low} \else M$_{\rm low}$\fi}
\def\aap{A\&A}

\def\apjl{ApJL}
\def\apjs{ApJS}

\newcommand{\oh}{\ifmmode 12 + \log({\rm O/H}) \else$12 + \log({\rm
O/H})$\fi}
\def\Oii{[O~{\sc ii}] $\lambda$3727}
\def\Oiii{[O~{\sc iii}] $\lambda\lambda$4959,5007}

\def\Oiiit{[O~{\sc iii}]$\lambda 4363$}
\newcommand{\Neiii}{[Ne~{\sc iii}] $\lambda$3869}
\newcommand{\Nev}{[Ne~{\sc v}] $\lambda$3427}

\def\flyf{\ifmmode f_{\rm Lyf} \else $f_{\rm Lyf}$\fi}
\def\pz{\ifmmode P(z) \else $P(z)$\fi}
\def\ki2{\ifmmode \chi^2 \else $\chi^2$\fi}
\def\zphot{\ifmmode z_{\rm phot} \else $z_{\rm phot}$\fi}

\newcommand{\xphot}{\ifmmode x_\gamma \else $v_\gamma$\fi}
\newcommand{\xobs}{\ifmmode x_{\rm obs} \else $x_{\rm obs}$\fi}
\newcommand{\xcmf}{\ifmmode x_{\rm CMF} \else $x_{\rm CMF}$\fi}
\newcommand{\vexp}{\ifmmode V_{\rm exp} \else $V_{\rm exp}$\fi}
\newcommand{\vmax}{\ifmmode V_{\rm max} \else $V_{\rm max}$\fi}
\newcommand{\nh}{\ifmmode N_{\rm HI} \else $N_{\rm HI}$\fi}
\newcommand{\dv}{\ifmmode \Delta v({\rm em-abs}) \else $\Delta v({\rm em}-{\rm abs})$\fi}

\def\fesc{\ifmmode f_{\rm esc} \else $f_{\rm esc}$\fi}
\def\fescrel{\ifmmode f_{\rm esc,rel} \else $f_{\rm esc,rel}$\fi}

\def\frellya{\ifmmode f^{\rm rel}_{\rm{Ly}\alpha} \else $f^{\rm rel}_{\rm{Ly}\alpha}$\fi}

\def\hii{H{\sc ii}}

\newcommand{\mstar}{\ifmmode M_\star \else $M_\star$\fi}
\newcommand{\muv}{\ifmmode M_{1500} \else $M_{1500}$\fi}
\newcommand{\auv}{\ifmmode A_{\rm UV} \else $A_{\rm UV}$\fi}
\newcommand{\luv}{\ifmmode L_{\rm UV} \else $L_{\rm UV}$\fi}
\newcommand{\lir}{\ifmmode L_{\rm IR} \else $L_{\rm IR}$\fi}
\newcommand{\lbol}{\ifmmode L_{\rm bol} \else $L_{\rm bol}$\fi}
\newcommand{\liruv}{\ifmmode L_{\rm IR+UV} \else $L_{\rm IR+UV}$\fi}
\newcommand{\liroveruv}{\ifmmode L_{\rm IR}/L_{\rm UV} \else $L_{\rm IR}/L_{\rm UV}$\fi}
\newcommand{\nlyc}{\ifmmode N_{\rm Lyc} \else $N_{\rm Lyc} $\fi}
\newcommand{\rholyc}{\ifmmode \rho_{\rm Lyc} \else $\rho_{\rm Lyc} $\fi}
\newcommand{\chion}{\ifmmode \xi_{\rm ion} \else $\xi_{\rm ion}$\fi}
\newcommand{\chioncorr}{\ifmmode \xi_{\rm ion}^0 \else $\xi_{\rm ion}^0$\fi}

\newcommand{\Civ}{C~{\sc iv}}
\newcommand{\Ciii}{C~{\sc iii}]}
\newcommand{\Ciiiuv}{C~{\sc iii}] $\lambda$1909}

\newcommand{\Niiiuv}{N~{\sc iii}] $\lambda$1750}
\newcommand{\Nivuv}{N~{\sc iv}] $\lambda$1486}

\newcommand{\source}{GN-z9p4}

\begin{document}

\title{Discovery of a new N-emitter in the epoch of reionization}
\subtitle{}
\author{D. Schaerer\inst{1,2}, 
R. Marques-Chaves\inst{1}, 
M. Xiao\inst{1},
D. Korber\inst{1}
}
  \institute{Observatoire de Gen\`eve, Universit\'e de Gen\`eve, Chemin Pegasi 51, 1290 Versoix, Switzerland
\and CNRS, IRAP, 14 Avenue E. Belin, 31400 Toulouse, France
}

\authorrunning{Schaerer et al.}
\titlerunning{Discovery of a new N-emitter at $z=9.4$ with JWST}

\date{Accepted for publication in A\&A Letters}


\abstract{We report the discovery of a compact star-forming galaxy at $z=9.380$ in the GOODS-North field (named \source) which shows numerous strong UV-optical emission lines and a single UV line, \Nivuv. This makes \source\ the third-highest redshift N-emitter known to date.
We determine the nebular abundances of H, C, N, O and Ne, size, and other physical properties of this object, and compare them to those of the other N-emitters known so far and  to other star-forming galaxies. Using the direct method we find a metallicity $\oh = 7.37 \pm 0.15$, one of the lowest among the N-emitters.
The N/O abundance ratio is highly super-solar, and C/O and Ne/O normal compared to other galaxies at low metallicity. 
We show that the compactness of \source\  (with effective radius $118\pm16$ pc at 2 \micron) and other N-emitters translates into very high stellar mass and SFR surface densities, which could be a criterium to identify other N-emitters. 
Future studies and larger samples are needed to understand these recently discovered, rare, and enigmatic objects.}


 \maketitle

\section{Introduction}
\label{s_intro}
Among the major surprises revealed by spectroscopic observations of the distant Universe with the James Webb Space Telescope (JWST) is arguably the spectrum of the $z=10.6$ galaxy GN-z11. It  shows a very peculiar rest-UV spectrum dominated by \Nivuv\ and \Niiiuv\ lines, which are usually not seen, and by \Ciiiuv\  \citep{Bunker2023JADES-NIRSpec-S}.
Quantitative analysis of the rest-UV and optical lines of this object showed quickly that the exceptional strength of the Nitrogen lines indicate a high (super-solar) N/O abundance in an object with low metallicity \citep[$\oh \approx 7.6-8.0$][]{Bunker2023JADES-NIRSpec-S,Cameron2023Nitrogen-enhanc,Senchyna2024GN-z11-in-Conte}, which was essentially unseen of until then.

The peculiar abundances of GN-z11 have triggered numerous studies exploring the nucleosynthetic origin and physical processes which can explain its special abundance ratios. A similarity with Globular Cluster abundances has been noted \citep[see][]{Senchyna2024GN-z11-in-Conte,Charbonnel2023N-enhancement-i}, and as of now, the following sources or scenarios have  been proposed to explain the observed high N/O abundance of GN-z11: Wolf-Rayet Stars, Very massive stars ($\sim 100-400$ \msun), AGB stars, Supermassive stars (with masses $\ga 10^{3-4}$ \msun\  and of normal metallicity or Population III), a top-heavy IMF, Wolf-Rayet stars with intermittent star-formation, or tidal disruption events \citep{Bunker2023JADES-NIRSpec-S,Cameron2023Nitrogen-enhanc,Charbonnel2023N-enhancement-i,Nagele2023Multiple-Channe,Watanabe2024EMPRESS.-XIII.-,Kobayashi2024Rapid-Chemical-,Bekki2023A-model-for-GN-,DaAntona-F.2023GN-z11:-Witness,Vink2023Very-massive-st,Maiolino2024A-small-and-vig,Nandal2024Explaining-the-}. No concensus has yet been reached on this question.
However, the possibility that N-emitters could be signposts of Globular clusters in formation, supermassive stars, or other ``exotic'' phenomena, clearly shows the interest of better understanding such objects.

After the discovery of GN-z11, several groups have searched for more objects exhibiting \Nivuv\ and/or \Niiiuv\ emission lines (hereafter referred to as N-emitters) resulting in 10 such objects, some of them previously known: Mrk 996, the Sunburst cluster, the Lynx arc, SMACS 2031, GS\_3073, RXCJ2248-4431, GLASS\_150008, CEERS-1019, GN-z11, and GHZ2/GLASS-z12
\citep[see][and references therein]{Senchyna2024GN-z11-in-Conte,Pascale2023Nitrogen-enrich,Marques-Chaves2024Extreme-N-emitt,Ji2024GA-NIFS:-An-ext,Topping2024Metal-poor-star,Isobe2023JWST-Identifica,Castellano2024JWST-NIRSpec-Sp}.
These objects include the core of a peculiar low-redshift dwarf galaxy (Mrk 996),  strongly lensed clusters or star-forming regions at $z \sim 2-4$, compact high-redshift galaxies from $z \sim 6-12$, and one Type 1 AGN at $z=5.55$. 
Among this diversity, all of them share super-solar N/O abundances, the presence of gas at unusually high densities, and they are found at metallicities $\oh \sim 7.4-8.3$.  

Here we report the discovery of a new N-emitter at $z=9.4$ (named \source), the third highest in redshift, whose rest-UV spectrum is dominated by \Nivuv. The available JWST observations allow us to determine accurate chemical abundances of H, N,O, Ne, and an upper limit on C. 
The source is among the, or possibly the most-metal poor N-emitter found so far. It is also very compact, suggesting that N-emitters may be exclusively found in regions of extremely high stellar mass and SFR densities. 


\begin{figure*}[htb]
\centering
\includegraphics[width=1\textwidth]{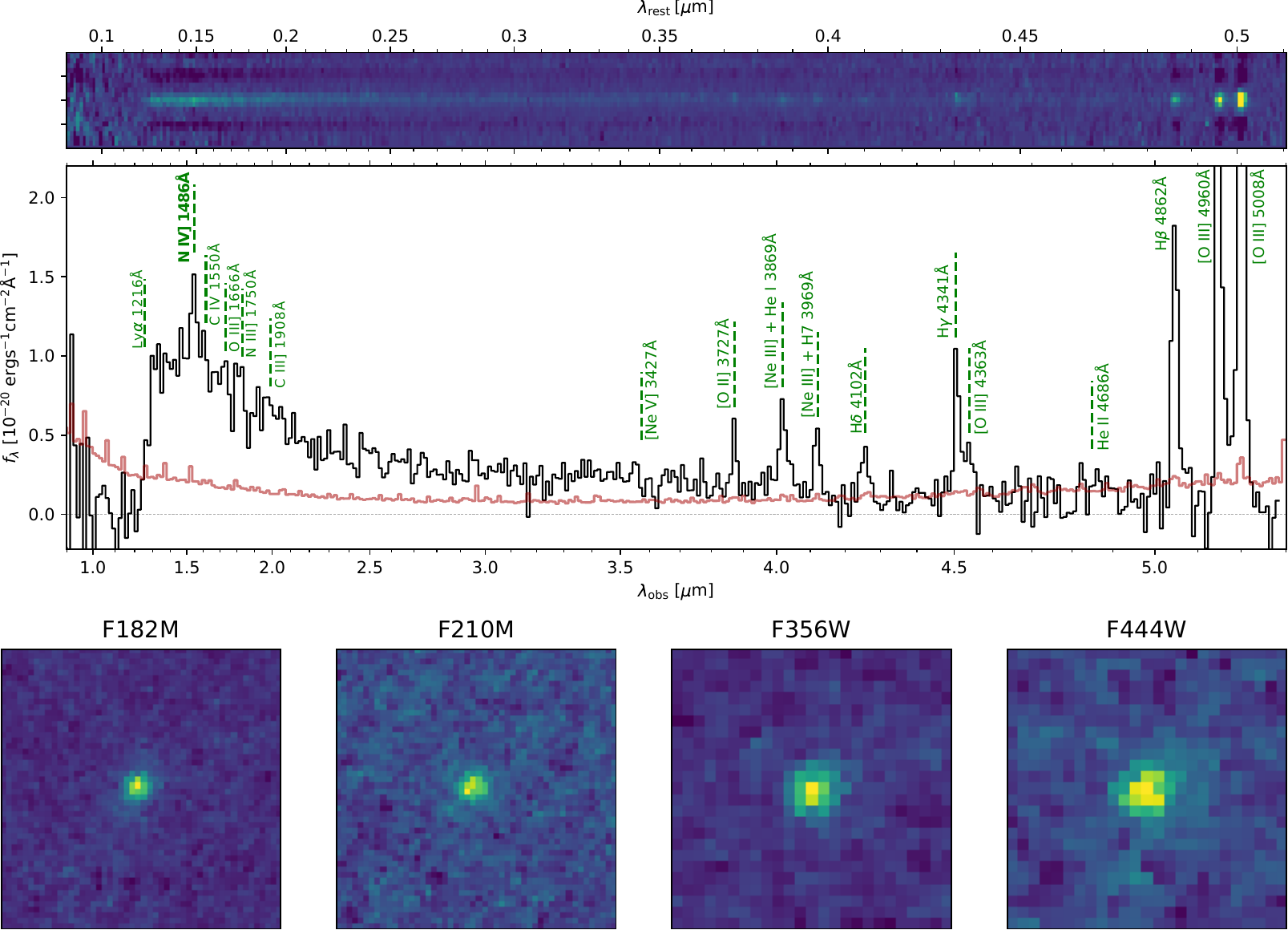}
\caption{{\em Top:} JWST NIRSpec/PRISM 1D and 2D spectra of  \source\ at $z=9.38$.  The low-resolution spectrum is shown in black, and the 1$\sigma$ uncertainty in red. Vertical dashed lines (green) mark the position of detected or un-detected nebular emission lines. X-axis at the bottom and top refer to the observed and rest-frame wavelengths in $\mu$m, respectively.  {\em Bottom:}  Postage stamps, of total size 1\arcsec x1\arcsec\ (corresponding 4.4 kpc x4.4 kpc physical size at $z=9.38$), of \source\ between 1.8 and 4.4 \micron. }
\label{fig_spec}
\end{figure*}

\section{\source: a new N-emitter at high redshift}
\label{s_obs}

\source\ ($\alpha$,~$\delta$ [J2000] = 189.016995, 62.241582) was identified as a $z \sim 9.4$ galaxy by one of the authors from searching the
DAWN JWST Archive (DJA)\footnote{\url{https://dawn-cph.github.io/dja}} database for distant galaxies. It was identified as an LBG earlier with a photometric redshift of $z = 9.5 \pm 0.4$ by \cite{Oesch2014The-Most-Lumino}. Inspection of the available JWST/NIRSpec PRISM spectrum quickly revealed a peculiar rest-UV spectrum dominated by one line, which we identified later as \Nivuv. 

\subsection{JWST NIRSpec and NIRCam observations}

 The JWST/NIRSpec spectrum of \source\ was taken by the  GTO program "NIRCam-NIRSpec galaxy assembly survey - GOODS-N"  on February 7, 2023 with the low-resolution PRISM ($R \simeq 100$) and coverage $\sim 1-5 \mu$m. The observations consist of slitlets of 3 shutters, with the 3-point nod with an aperture position angle $\rm PA \simeq 19.58$ deg (see details in \citealt{Eisenstein2023Overview-of-the}). The total exposure time was 6127 seconds, split into 6 individual exposures of 14 groups each. 
The reduced spectrum,  shown in Figure  \ref{fig_spec}, was obtained from the DJA  database. The calibration reference data system context jwst\_1183.pmap was used to correct spectra for flat-field and implement the wavelength and flux calibrations. 
Details of the data reduction are given in \cite{Heintz2024The-JWST-PRIMAL}.  

\source\ was also observed with JWST/NIRCam with the F182M, F210M, F356W, and F444W filters. These images are obtained from the DJA Imaging Data Products and were reduced using the grizli reduction pipeline, which includes procedures for masking the snowball artifacts and minimizing the impact of 1/f noise. 
For the photometry we proceed as in \cite{Weibel2024Galaxy-Build-up}.
\source\ shows a very compact morphology in all NIRCam filters, as will be quantified below.

To account for wavelength-dependent slit losses and absolute flux calibration, we derive the synthetic photometry of NIRSpec spectrum through each NIRCam filter bandpass and compared it to that obtained from observed photometry. We find that the synthetic flux densities match well, with an average offset of $\simeq 16\%$, within the 1$\sigma$ uncertainties of the PRISM spectrum. No correction was thus made.

\begin{table}[htb]
\begin{center}
\caption{Emission line flux (in $10^{-18}$ erg s$^{-1}$ cm$^{-2}$) and equivalent width (in \AA) measurements or $3 \sigma$  upper limits  for \source. 
\label{tab_flux_measurements}}
\begin{tabular}{l c c }
\hline \hline
\smallskip
\smallskip
Line & Flux  & EW$_{\rm rest}$\\
\hline 
N~{\sc iv}] $\lambda\lambda  1483, 1486$  & $2.913\pm 0.53$ & $31.8\pm   6.46$ \\
C~{\sc iv} $\lambda\lambda 1548,1550$ & $ < 1.56$  & $<19.1$ \\
{O~{\sc iii}]} $\lambda\lambda 1661,1666$ &  $<1.08$ & $<14.7$ \\
{N~{\sc iii}]} $\lambda 1750$ & $<1.32$ & $<19.7$ \\
C~{\sc iii}] $\lambda\lambda 1907,1909$ & $<1.11$ & $<18.9$ \\
{[Ne~{\sc v}]} $\lambda 3427 $ & $<0.39$ & $<15.9$ \\
{[O~{\sc ii}]} $\lambda\lambda 3727,3729$ & $0.683\pm 0.20$ & $38.8\pm  13.57$ \\
{[Ne~{\sc iii}]}+HeI $\lambda 3869$ & $1.374\pm 0.24$ & $84.9\pm  24.08$ \\
{[Ne~{\sc iii}]} $+$ H7 $\lambda 3968$ & $0.824\pm 0.15$ & $54.0\pm  17.4 $ \\
H$\delta$ $\lambda 4101$ & $0.976\pm 0.223$ & $111.7\pm  46.8$ \\
H$\gamma$ $\lambda 4340$ & $1.760\pm 0.18$ &$151.0\pm  22.0 $ \\
{[O~{\sc iii}]} $\lambda 4363$ & $0.691\pm 0.16$ & $59.3\pm  15.3 $ \\
\Heiiopt\ & $<0.54 $ & $<44.7$ \\
H$\beta$ $\lambda 4861$ & $3.314\pm 0.14$ &$284.4\pm  34.0 $ \\
{[O~{\sc iii}]} $\lambda 4959$ & $6.032\pm 0.24$ &$517.6\pm  42.7 $ \\
{[O~{\sc iii}]} $\lambda 5007$ &  $16.250\pm 0.18$ &$1394.5\pm 112.2 $ \\
\hline 
\end{tabular}
\end{center}
\end{table}

\subsection{Emission line measurements}

As shown in Figure \ref{fig_spec}, \source\ shows a blue spectrum, strong emission lines in the rest-frame optical, and a single clear emission line detection in the rest-UV. The systemic redshift of \source\ derived from the rest-optical lines (centroids of \hb, \Oiii) is $z_{\rm sys}=9.37956 \pm 0.00017$. The rest-UV line is therefore clearly identified as \Nivuv. Since this is the strongest rest-UV line,  we classify this galaxy as a N-emitter, in analogy with the other  rare objects recently discovered with the JWST (see above).
After GLASS-z12 at $z=12.342$ and GN-z11 at $z=10.6$, this detection makes \source\ the third most distant galaxy among the class of rare N-emitters, which now counts 6 objects at $z>6$ \citep[see][]{Castellano2024JWST-NIRSpec-Sp,Bunker2023JADES-NIRSpec-S,Marques-Chaves2024Extreme-N-emitt,Topping2024Metal-poor-star,Isobe2023JWST-Identifica} and 11 in total.

 The rest-optical spectrum of \source\ shows H lines from the Balmer series, \Oii, \Oiii, the auroral \Oiiit, and \Neiii+\hei, which have now often been seen in JWST spectra. In total, we detect 10 emission lines with a significance larger than 3 $\sigma$.  To measure the line fluxes we  fit simultaneously Gaussian profiles and a power-law to the continuum, as described in \cite{Marques-Chaves2024Extreme-N-emitt}. Table \ref{tab_flux_measurements} summarizes our flux and rest-frame equivalent width measurements, as well as upper limits for several important lines.
We note that \source\ shows very strong lines, with equivalent widths exceeding largely those of the previously identified N-emitters at $z>8$ (CEERS-1019, GN-z11, GLASS-z12).
The Balmer lines are compatible with no extinction, although with large uncertainties, allowing $0 \le$ E(B-V) $\la 0.2$.

\section{Physical properties of \source\ and other N-emitters}
\label{s_results}

\subsection{Nature of \source }
\label{s_nature}
What powers the emission lines of \source\ ?
%
None of the high-ionization lines observed in AGN are detected. For \heii\ we derive a $3\sigma$ upper limit He~{\sc ii}/H$\beta \leq 0.16$, and 
\Nev\ is also not detected with \Nev/\Neiii\ $ \la 0.28$. 
Both of these limits are close to the SF/AGN boundary or in the ``composite'' range and compatible with photoionization by stars \citep{Shirazi2012Strongly-star-f,Cleri2023Using-Ne-V/Ne-I}. Classical optical diagnostics (the so-called BPT diagrams) and rest-UV emission line diagnostics are not available for \source, since the required lines are not detected or not covered by the NIRSpec observations.
In fact, all observed emission line ratios are typical of compact, metal-poor star-forming galaxies, as can be seen by a simple comparison with 
 \cite{Schaerer2022First-look-with} showing SDSS measurements from \cite{Izotov2021Low-redshift-co}.
Finally, the spectrum of source shows no broad lines, although the resolution is low.
More precisely, we measure a FWHM$(\hb) = 434 \pm 65$ km/s which is basically unresolved, and we do not detect any significant broad component in \hb, \Oiii, or other lines. This excludes a substantial contribution of a type 1 AGN (FWHM$>1000$ km/s); e.g., if a type-1 AGN contributes $>50$\% of the \hb\ flux, we would expect FWHM(\hb) $> 700$ km/s.
In short, from the available data we find no indication for AGN, and the emission lines are compatible with those of metal-poor star-forming galaxies, although better spectra may be needed to establish this more firmly.

\begin{table}[htb]
\caption{ISM properties, ionic and total heavy element abundances (derived assuming no extinction) and other properties of \source}
\label{ta_abund}
\begin{center}
\begin{tabular}{lrrrrr}
\hline \hline
Property & Quantity \\ \hline 
$T_{\rm e}$(O {\sc iii}) [K] & $23018 \pm 3727$\\
$T_{\rm e}$(O {\sc ii}) [K] &  $14685 \pm 1328$\\
12+log(O$^{+}$/H$^+$) &  $6.30 \pm 0.22$ \\ 
12+log(O$^{2+}$/H$^+$) & $7.36 \pm 0.15$ \\
12+log(O/H)  &           $7.37 \pm 0.15$ \\
log(N$^{3+}$/H)   & $-5.22 \pm 0.38$  \\
log(N$^{3+}$/O)   & $-0.59 \pm 0.24$  \\
log((C$^{2+}$+ C$^{3+}$)/O)  & $< -1.18$  \\
ICF(Ne$^{2+}$/O$^{2+}$) & 1.04 \\
log(Ne/O)  &  $-0.75 \pm 0.08$ \\
\\
$r_{\rm eff}$(F182M and F210M) & $118 \pm 16$ pc \\
$r_{\rm eff}$(F356W and F444W) & $<190 $ pc \\
\muv & $-21.0$  \\
$\log($\mstar$)$ & $8.7 \pm 0.2$ \msun \\ 
SFR  & $64 \pm 14$ \msunyr \\
sSFR & $120 \pm 40$ Gyr$^{-1}$ \\
E(B-V) & $0.1 \pm 0.05$ \\
\hline 
\vspace*{-1cm}
\end{tabular}
\end{center}
\end{table}%

\begin{figure}[tb]
\centering
\includegraphics[width=0.5\textwidth]{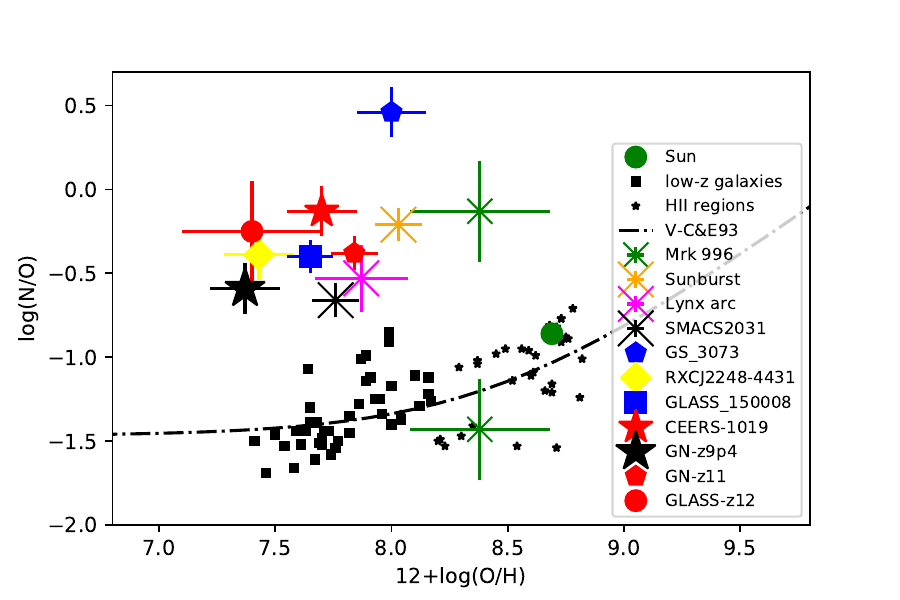}
\includegraphics[width=0.5\textwidth]{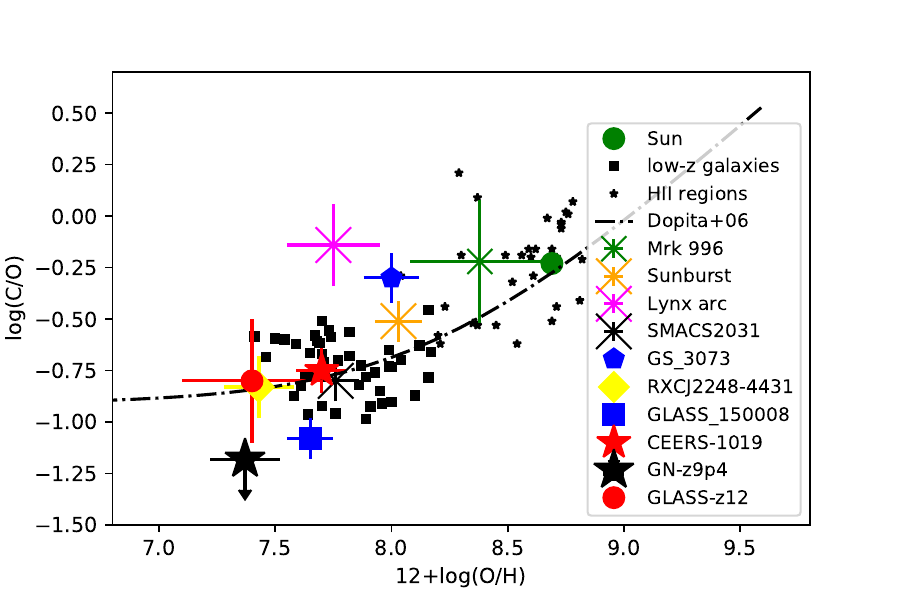}
\caption{Observed chemical abundances of all known N-emitters including \source. {\em Top:} N/O versus O/H, {\em bottom:} C/O versus O/H.
\source\ is shown by a black star. The N-emitters are sorted by increasing redshift in the legend. Only one measurement (for the densest region) is shown for GS\_3073. For Mrk 996, two N/O are available, probing regions of low and high density \citep{James2009A-VLT-VIMOS-stu}.
Low-$z$ star-forming galaxies and \hii\ regions from the compilation of \cite{Izotov2023Abundances-of-C} are shown by small black symbols.
The dash-dotted line shows the average trend observed in low-$z$ star-forming galaxies, as parametrized by \cite{Vila-Costas1993The-nitrogen-to} for N/O and C/O by \cite{Dopita2006Modeling-the-Pa} respectively.}
\label{fig_abund}
\end{figure}

\subsection{Ionic and total metal abundances}
\label{s_abund}
With a 4.3 $\sigma$ detection of the auroral \Oiiit\ line, it is possible to determine accurate abundances of the ISM in \source, using the direct method.
The electron temperature $T_{\rm e}$(O~{\sc iii}) is used to obtain abundances of ions O$^{2+}$, N$^{3+}$,  Ne$^{2+}$, and upper limits for 
C$^{3+}$ and C$^{2+}$; the temperature in the low-ionization region, $T_{\rm e}$(O~{\sc ii}), to derive the ionic abundance of O$^+$.
In the absence of direct constraints on the electron density, we adopt $n_e=100$ \cmc\ (low density regime) 
and for densities $n_e \la 10^5$ \cmc\ our results are essentially unchanged (abundance ratios approximately within the quoted uncertainties).
Ionic abundances are derived following \cite{Izotov2006The-chemical-co} for the optical lines. 
For Ne$^{2+}$ we use the ionization correction factor (ICF) following \citet{Izotov2006The-chemical-co}. 
The  N$^{3+}$/H$^+$ abundance is determined from the N~{\sc iv}]/\hb\ ratio, and upper limits on the Carbon abundance from the  \Ciii\ and \Civ\ intensities with respect to \hb, following \cite{Villar-Martin2004Nebular-and-ste}.  
The results are listed in Table \ref{ta_abund} (assuming no extinction), and the CNO abundances are shown and compared to other objects in Fig.~\ref{fig_abund}.
Adopting the median extinction from SED fits does not alter O/H, but would imply an increase of N/O (the C/O limit) by a factor $\sim 2$ (1.6). 

We derive a total oxygen abundance of $\oh = 7.37 \pm 0.15$, approximately 5\% solar \citep[assuming a solar value of \oh=8.69;][]{Asplund2009The-Chemical-Co}, which is dominated by the ionic abundance of O$^{2+}$/H$^+$. This makes \source\  the N-emitter with the lowest-metallicity determined from the direct method.
The ionic abundance of N$^{3+}$ indicates a very high N/O abundance ($\log($N$^{3+}$/O)$-0.59 \pm 0.24$), approximately 1.9 times solar,  for the low metallicity \citep[cf.][]{Asplund2021The-chemical-ma}.
From the non-detection of the Carbon lines we derive an upper limit of $\log$(C/O)$<-1.18$, sub-solar but fairly representative of C/O in other metal-poor galaxies \citep{Izotov2023Abundances-of-C}.
Finally, the Neon abundance ratio $\log({\rm Ne/O}) = -0.75 \pm 0.08$ is compatible with the average value of $\log({\rm Ne/O}) = -0.80 \pm 0.01$ determined for normal star-forming galaxies by \cite{Guseva2011VLT-spectroscop} at the same metallicity.

Together with the strongly lensed star-forming clump RXCJ2248-4431 at $z=6.1$ where the metallicity has also been measured from the direct method \citep{Topping2024Metal-poor-star}, and with GHZ2 at $z=12.34$ \citep{Castellano2024JWST-NIRSpec-Sp,Zavala2024Detection-of-io}  where the metallicity is inferred only indirectly, \source\ is among the lowest-metallicity N-emitters known so far. 

Taken together, the derived abundances of \source\ show that this object has a ``metallicity'' (O/H) of approximately 5\% solar, an very high N/O abundance, and a fairly normal C/O abundance, when compared to galaxies of similar metallicity (see Fig.\ \ref{fig_abund}). 
These unusual abundance patterns are shared with the other known N-emitters, which are also shown in color in this figure. 
The origin of the observed N-enrichment and other abundance ratios has been discussed earlier (see Introduction), and is subject of intense research. It is beyond the scope of this Letter.

For very high electron densities, $n_e > 10^5$ \cmc, the inferred electron temperature would be lower, leading to a higher O/H and even more extreme (higher) N/O and C/O ratios, further highlighting the peculiarity of this source.

\begin{figure}[tb]
\centering
\includegraphics[width=0.5\textwidth]{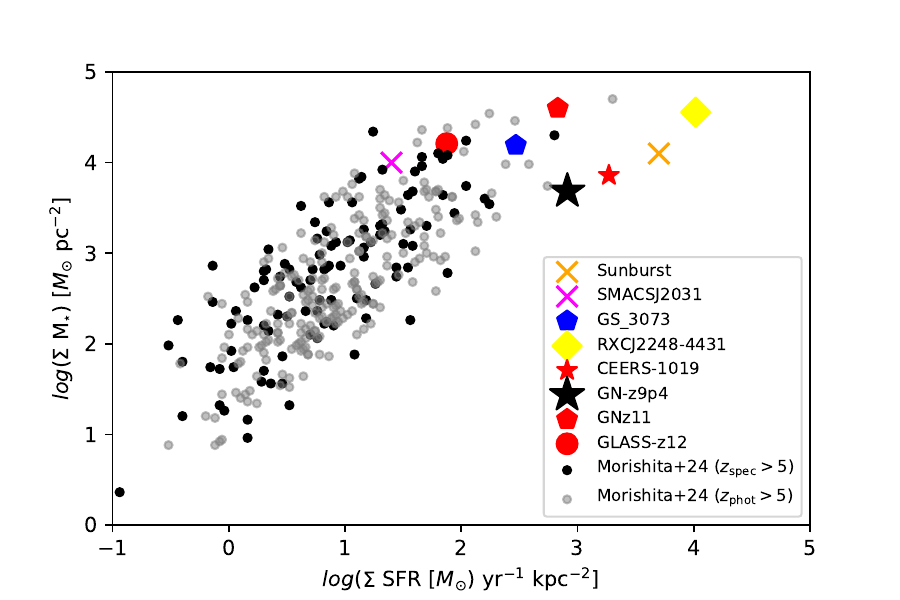}
\caption{Observed mass and SFR surface densities of the known N-emitters including \source\ compared to the sample of 341 star-forming galaxies at $5 < z< 14$ observed with JWST, analysed by  \cite{Morishita2024Enhanced-Subkil}. Among those, objects with spectroscopic and photometric redshifts are distinguished.}
\label{fig_sigma}
\end{figure}

\subsection{Compactness, mass and SFR surface density of \source}

 We investigate the morphology of \source\ using the PySersic code \citep{Pasha2023pysersic:-A-Pyt} to fit its light distribution in the four NIRCam filters. This process uses a 2D Gaussian function convolved to the instrumental point spread function (PSF) obtained from nearby bright stars in the NIRCam field-of-view.  The fitting process is performed on $\simeq 2^{\prime \prime} \times 2^{\prime \prime}$ background-subtracted cutouts centered on \source. To check whether the source is spatially resolved we compare the residuals obtained from the best-fit model using a Gaussian profile to that obtained with a point source. We find that \source\ is compact, but resolved in the NIRCam F182M and F210M bands, for which we obtained a consistent effective radius of $r_{\rm eff} = 118 \pm 16$ pc. At longer wavelengths, the best-fit models assuming Gaussian and point source profiles yield both similar residuals, suggesting that \source\ is unresolved in F356W and F444W ($r_{\rm eff} < 190$ pc). 

To determine the main properties of the stellar content of \source\ we have used the CIGALE and Bagpipes codes of \cite{Boquien-M.2019CIGALE:-a-pytho,Carnall2018Inferring-the-s}  to fit the multi-band SED, and we have added fluxes of the main emission lines and/or the full PRISM spectrum in the fits. Redshift and metallicity are fixed and a variety of star-formation histories and attenuation laws were explored\footnote{We adopt a concordance cosmology with $\Omega_{\rm m} = 0.272$, $\Omega_{\rm \Lambda} = 0.728$, and $H_{0} = 70.4$ \kms\ Mpc$^{-1}$. We have explored exponentially declining and delayed star-formation histories, Calzetti and SMC attenuation/extinction law.}. 
The overall results do not strongly depend on these assumptions, in particular since the spectrum is largely dominated by a young stellar population from the rest-UV to 5.4 \micron. For this reason the stellar mass is, however, fairly uncertain. All SED fits indicate the presence of some dust attenuation. The main derived quantities, obtained using the SMC extinction law, are summarised in Table \ref{ta_abund}.

\subsection{N-emitters are compact and rare}
As already noticed early on, N-emitters show high ISM densities ($n_e \sim 10^{3-5}$ \cmc), or indications for regions with different densities, including densities spanning a wide range, up to high densities \citep[see][]{Patricio2016A-young-star-fo,Ji2024GA-NIFS:-An-ext}.
No density diagnostic is available from the PRISM spectrum of \source, but we suspect that it also contains some gas at high electron densities.

Another property shared by N-emitters is their compactness, as illustrated in Fig.~\ref{fig_sigma}, where we show the stellar mass and SFR surface density of these objects, compared to those of star-forming galaxies at $5 < z <14$ measured by \cite{Morishita2024Enhanced-Subkil} using JWST observations.
For the N-emitters its is important to note that these quantities were not determined in a uniform fashion, since, for example, different assumptions were made for the SED  fits to derive stellar masses, and most importantly the effective spatial resolution is much higher for lensed objects. Despite this, 
it is clear that the N-emitters are exclusively found among the sources with the highest stellar mass and SFR surface densities, which indicates that the observed strong N-enhancement is related to or found in the most compact star-forming regions or objects. 
If confirmed with larger statistical samples, this suggests that the source of N-enhancement or the  physical processes leading to a high N/O abundance require very high mass and/or SFR surface densities. This, and the high ISM densities observed in N-emitters, could represent interesting constraints to distinguish different  scenarios to explain their origin.

Interestingly, several other objects from the JWST sample analysed by \cite{Morishita2024Enhanced-Subkil}  occupy the same area as the known N-emitters and are also covered by spectroscopic observations. We have therefore examined the available NIRSpec PRISM spectra, finding one or two possible new N-emitters, although the quality of the spectra is limited.
This indicates that a selection by stellar mass and SFR surface density could be an efficient way to find new N-emitters.

The finding of $\sim 2$ N-emitters among the 109 galaxies of \cite{Morishita2024Enhanced-Subkil}  having JWST spectra, shows that these objects are rare, even at high-redshift  ($z>5$). This is also clear from the fact that the Nitrogen lines in the UV (\Nivuv, \Niiiuv) are not detected in stacked NIRSpec spectra of $z >5$ galaxies \citep{Roberts-Borsani2024Between-the-Ext,Langeroodi2024NIRSpec-View-of}.  
These current data indicate that approximately 7 out of 500 ($\sim 1-2$\%) galaxies spectroscopically covered by JWST are N-emitters. This simple estimate should, however, be taken with a grain of salt, since no systematic search has yet (to the best of our knowledge) been undertaken for these sources,  the target selection function is not well known, and the depth and signal-to-noise of the various observations vary strongly. Proper statistical studies of the occurrence of the N-emitters remain therefore to be done.

\section{Conclusions}
\label{s_conclude}

Examining JWST/NIRSpec PRISM observations  we have identified a compact star-forming galaxy at $z=9.436$ in the GOODS-North field which shows numerous strong UV-optical emission lines and a single UV line, \Nivuv, which qualifies this object as a new N-emitter. 
This brings the total number of these rare and enigmatic objects to 11. 

From the emission lines, including the auroral \Oiiit\ line, we have determined the abundances of H, N, O, Ne, and an upper limit on C, finding $\oh = 7.37 \pm 0.15$, N/O$=-0.59 \pm 0.24$, Ne/O$=-0.75 \pm 0.08$, and C/O$<-1.18$.  These properties make \source\ the third highest-redshift (after GLASS-z12 and GN-z11) and one of the most metal-poor N-emitters (with RXCJ2248 and possibly GLASS-z12) known so far.

With a super-solar N/O abundance ratio and a fairly normal C/O for low-metallicity (O/H),  the observed ISM abundances of \source\ are found to be similar to most of the previously identified N-emitters, whose origin is debated. 
Comparing the stellar mass and SFR surface densities of the known N-emitters with those of star-forming galaxies, we have shown that N-emitters are exclusively found at the high-end tail of the distribution (typically with $\log(\Sigma_{\mstar}) \ga 3.5$ \msun pc$^{-2}$ and $\log(\Sigma_{\rm SFR}) \ga 2$ \msun yr$^{-1}$ kpc$^{-2}$), indicating that this phenomenon probably requires peculiar conditions (e.g.~high ISM densities and compact regions), physical processes or ``exotic'' sources of nucleosynthesis. 

Finally, we estimate that approximately $\sim 1-2$ \% of the galaxies currently observed with JWST at $z \ga 5$ are N-emitters, quantifying that these are rare objects. We also speculate that this phenomenon could be more frequent at higher redshift.


\bibliographystyle{aa}
\bibliography{merge_misc_highz_literature.bib}
\end{document}